# Link between cascade transitions and correlated Chern insulators in magic-angle twisted bilayer graphene

Qianying Hu[1,2,#], Shu Liang[3,#], Xinheng Li[2], Hao Shi[3], Xi Dai[3,*], Yang Xu[1,2,*]

[1]Beijing National Laboratory for Condensed Matter Physics, Institute of Physics, Chinese Academy of Sciences, Beijing, China
[2]School of Physical Sciences, University of Chinese Academy of Sciences, Beijing, China
[3]Department of Physics, Hong Kong University of Science and Technology, Kowloon, Hong Kong, China

[#]These authors contributed equally to this work.
[*]Email: daix@ust.hk, yang.xu@iphy.ac.cn

**Chern insulators are topologically non-trivial states of matter characterized by incompressible bulk and chiral edge states. Incorporating topological Chern bands with strong electronic correlations provides a versatile playground for studying emergent quantum phenomena. In this study, we resolve the correlated Chern insulators (CCIs) in magic-angle twisted bilayer graphene (MATBG) through Rydberg exciton sensing spectroscopy, and unveil their direct link with the zero-field cascade features in the electronic compressibility. The compressibility minima in the cascade are found to deviate substantially from nearby integer fillings (by $\Delta v$) and coincide with the onsets of CCIs in doping densities, yielding a quasi-universal relation $B_c=\Phi_0\Delta v/C$ (onset magnetic field $B_c$, magnetic flux quantum $\Phi_0$ and Chern number $C$). We suggest these onsets lie on the intersection where the integer filling of localized "$f$-orbitals" and Chern bands are simultaneously reached. Our findings update the field-dependent phase diagram of MATBG and directly support the topological heavy fermion model.**

Two-dimensional electronic systems hosting nontrivial Chern bands can give rise to the quantum anomalous Hall effect (QAHE) with spontaneous breaking of the time-reversal symmetry at zero magnetic fields (*1–10*). However, in many systems such as the magnetic topological insulator MnBi$_2$Te$_4$ (*11, 12*), MATBG (*13–22*), and multilayer crystalline graphene (*23–25*), the Chern insulators ($C\neq0$) often need to be stabilized by a finite magnetic field. Understanding the driving force of this finite field is crucial for establishing unified theoretical frameworks. MATBG with multiple robust CCIs provides a good platform, where the CCIs in MATBG have been found to emanate from integer fillings $v$ of the moiré superlattice and follow the regular sequence of $(C, v)$ = ($\pm3, \pm1$), ($\pm2, \pm2$), and ($\pm1, \pm3$) (*13–22*).

The nature of the ordered ground state at low temperatures is often closely related to its less ordered normal state at high temperatures, as exemplified by the link between superconductivity and strange metallicity in cuprates (*26*, *27*). In MATBG, the interplay between spin, valley, and sublattice degrees of freedom, coupled with the helical Dirac electrons and narrow moiré bandwidth, creates a diverse ensemble of competing many-body ground states (*28–31*). The normal states for the low-temperature correlated insulators and superconductivity are suggested to be the so-called "cascade transitions", manifested as asymmetric sawtooth features commonly observed in the doping ($n$) dependences of the thermodynamic compressibility ($\partial n/\partial \mu$) or local spectroscopic measurements (*20*, *28*, *29*, *32–36*). However, no consensus has been reached for these states. They transit to the CCIs under finite magnetic fields, while what governs the energy hierarchy and the critical magnetic fields ($B_c$) for these CCIs is poorly understood.

Here we report studies of twisted bilayer graphene (TBG) samples with $\theta$=1.09°~1.25° close to the magic angle through Rydberg exciton sensing (*37–39*). The variation of thermodynamic compressibility of TBG with the filling factor $v$ and magnetic field $B$ can be well captured. By carefully analyzing the results, we establish a straightforward link between the cascade transitions and CCIs, i.e., the onset of the CCIs always aligns with the doping density where the electronic compressibility is minimized at zero magnetic fields. In other words, the ratio between $\Delta v$ and $B_c$ equals a constant $C/\Phi_0$. This observation extends across all six CCIs in MATBG and is discernible over a broader range of twist angles. The understanding of such experimental observations can be unified by the topological heavy fermion (THF) model. In AA-stacked sites, the integer filling of localized orbitals, known as "*f*-electrons", leads to the suppressed quasi-particle weight and enhanced local moments. Meanwhile, the remaining electrons, differentiated by $\Delta v$ and referred to as "*c* electrons", occupy AB-stacked sites from topological conduction bands and are responsible for the appearance of CCIs under magnetic fields. Our findings elucidate the intrinsic driving mechanism of the finite $B_c$, provide experimental evidence supporting the THF model (*36*, *40–42*), and offer a new perspective on the physical origin of the cascade transitions.

**Sawtooth feature at $B = 0$ and deviation from integer fillings**

The device schematic is shown in Fig. 1A. A monolayer $WSe_2$ sensor is placed adjacent to near-magic-angle TBG and encapsulated by hBN/graphite dielectrics where a gate voltage $V_g$ is applied to tune the carrier density $n$ in TBG. The optical detection has a spatial resolution of about 1 μm. In Fig. 1B, the doping-dependent reflectance contrast spectra of a 1.09° TBG/$WSe_2$ device are measured at a temperature $T = 1.7$ K. The TBG barely has optical responses in this energy range and the main resonances arise from the $WSe_2$ excitons. The 1s excitonic state (~1.70 eV) of the monolayer $WSe_2$ exhibits small variations while the 2s resonance (~1.78 eV) features a sawtooth pattern ($|n|<3\times10^{12}$ cm$^{-2}$), whose energy shift and intensity reflect the dynamic evolution of the electronic compressibility of TBG (refer to Refs. (*37–39*) for details). For clarity, we present the color map of this region with its energy subtracted by a slowly varying smooth background as $\Delta E$ in Fig. 1C. The doping density is converted to the filling factor $v$, the number of electrons ($v > 0$) or holes ($v < 0$) per moiré unit cell. The comprehensive data analyzing procedure is outlined in the Methods, and additional details can be found in Fig. S1.

The thermodynamic ground state of the electronic phases in MATBG is mainly characterized by its compressibility $\partial n/\partial \mu$, which can be accessed by several experimental approaches including quantum capacitance (22), scanning electron transistor (16, 20, 32), and direct chemical potential ($\mu$) measurements (34). Indeed, the $\Delta E$ of the 2s state distinctly mirrors the asymmetric sawtooth behavior in the direct measurement of inverse compressibility ($\partial \mu/\partial n$) of MATBG, thereby affirming the validity of our method. When MATBG enters the compressibility minimum, the dielectric screening effect on the 2s exciton weakens, resulting in a peak in energy and enhanced spectral intensity for the 2s exciton (39, 43). The zero-field sawtooth feature in the compressibility was initially interpreted as a cascade of phase transitions where the flavor (spin or valley) symmetry is spontaneously broken near the integer fillings (19, 32). However, direct evidence for symmetry breaking in the zero-field cascade transitions is lacking, and this concept is at odds with the observations that these states persist up to tens of Kelvin — a temperature comparable to the moiré bandwidth of approximately 10 meV (44, 45).

Notably, in addition to the well-defined peaks at the $v = 0$ charge-neutral point and the full-filling moiré superlattice gaps at $v = \pm 4$, all six supplementary peaks around $v = \pm 1, \pm 2, \pm 3$ deviate toward higher density from nearby integer filling factors as indicated by the dashed lines and red bars. This deviation is also observed in many previous studies but has not yet been thoroughly analyzed or discussed (20, 22, 32, 34). We quantitively extract the deviation $\Delta v$ of compressibility minima from nearby integers in Fig. 1C. The values adapted from previous global (22, 34) and local (32) compressibility measurements on MATBG are also shown in Fig. 1D. This deviation $\Delta v$ is discernible for all the six states, and is always positive on the electron side and negative on the hole side. Among different studies, the deviation in the hole doping side is more prominent than the electron doping side, whose absolute value can be as large as ~0.4 for the $v = -1$ state. The exact value shows slight variation between different studies, probably due to variations in the local strain or twist angles. This phenomenon extends beyond what could be attributed to experimental artifacts and may share the same origin with the resistance peak deviation from integers observed by several mesoscopic charge transport studies (21, 34, 35).

**Correlated Chern insulators (CCIs) under finite magnetic fields**

Following the zero-field measurements, we apply different perpendicular magnetic fields, and the spectral evolution is depicted in Fig. 2a. With increasing field strength, the single peak feature at $v = 0$ transforms into a cluster of peaks spanning a wider range in doping density, so are those at $v = \pm 4$. On the other hand, the peaks around $v = \pm 1, \pm 2, \pm 3$ become more defined after reaching a specific magnetic field strength and shift towards higher doping densities. In the top panel with $B = 9$ T, all the peaks are sharp in density and feature blue tails in energy, indicating the presence of incompressible gapped states.

As mentioned earlier, the evolution of compressibility in TBG manifests through two primary effects on the spectrum of Rydberg excitons: alterations in resonance energy and spectral intensity. To comprehensively examine the field-dependent evolution of these states, we extract these two values from the spectra (see details in Methods and Fig. S1) and construct the fan diagrams presented in Fig. 2B and 2C. The fan diagrams extracted from $\Delta E$ and intensity exhibit similarities and the topologically nontrivial gapped states originating from different integer moiré fillings are highlighted by the colored lines in Fig. 2D. They correspond to the integer quantum Hall states (IQH, guided by blue lines) with LL filling factors $v_{LL}$ and CCIs (guided by red lines) with Chern number $C$. Both $v_{LL}$ and $C$ can be determined from the Streda formula $\Phi_0 dn/dB$ according to their slopes in the fan diagram.

Emanating from the charge neutrality point, the primary gapped states correspond to the empty/filled zeroth LL with $v_{LL}=\pm 4$ and the trivial gaps at $v_{LL}=0$. All the other symmetry-broken quantum Hall ferromagnetic states at $0<|v_{LL}|<4$ are discernible at $B>\sim 2$ T. Different experiments have revealed different sequences (being four-fold, two-fold, or single-fold degenerate) of the LLs originating from $v=0$ (*13–22, 46*). It is generally believed that the energy gaps are the strongest at $v_{LL}=\pm 4$, $\pm 2$, and $0$, consistent with our observations. Additionally, two sets of fully degeneracy lifted LLs (with $v_{LL}= -4$ to $-1$ and $1$ to $4$, respectively) originate from the full filling gaps at $v = \pm 4$ after $B > \sim 4$ T, and only the fans towards higher densities are retained.

In addition, six states labeled in red evolve from $v = \pm 1, \pm 2, \pm 3$ and become visible after reaching a certain critical magnetic field $B_c$, signifying the CCIs. It is typically understood that due to strong electronic interactions, the MATBG favors sequential fillings of the topologically nontrivial Hofstadter subbands with Chern numbers $C =-1$ for the holes and $C =1$ for electrons (*47*). Each Hofstadter subband is flavor polarized and corresponds to a single hole or electron occupying per moiré unit cell. This model can give the right sequence of Chern number $C$ as a function of $v$ at the integer fillings, yielding $(C, v) = (\pm 3, \pm 1), (\pm 2, \pm 2)$, and $(\pm 1, \pm 3)$. However, the certain value of the $B$ field necessary to stabilize these CCIs in previous reports shows large variations without any universal pattern (*13, 14, 18–20, 22*).

Here, we find the critical magnetic field $B_c$ is determined by the deviation $\Delta v$ in the zero-field sawtooth feature. In Fig. 2D, the zero-field spectrum is placed beneath the fan diagram for reference. The peaks in the sawtooth feature consistently align with the CCI onsets. This relation can be seen in the fan diagram in Fig. 2b and 2c where the onsets of the CCIs coincide with the vertical broader features that extend to zero magnetic field (indicated with blue bars in Fig. 2d). These features correspond to the low-field compressibility minima in the fan diagram, manifested as enhanced energies and oscillator strengths in the exciton resonance. It could be numerically expressed as $B_c = \Phi_0 \Delta v/C$ obtained from the slope of the CCIs following $\Delta v/B_c=dn/dB=C/\Phi_0$. The observed straightforward relationship applies well to all the six CCIs within the flat band, which suggests that the two independent topics mentioned earlier in this paper (the deviation of compressibility minima from integers in the cascade transitions & the critical $B$ field of the CCIs) are strongly intertwined.

**Link between cascade transition and CCIs**

To understand the identified link between the compressibility minimum and the emergence of CCIs, we first look back at the cascade transitions and consider their origins. The deviation of the compressibility minima from the nearby integer has two interesting features. First, the deviation only happens at non-zero integer fillings (±1, ±2, ±3). There is no deviation at all at the charge neutrality point (zero filling). Second, the deviation is positive on the electron side and is negative on the hole side, showing some approximate "particle-hole symmetry". Both two features can be well explained by the THF model for MATBG, where the flat bands can be decomposed into two different orbitals, the "$f$-orbitals" located at the AA stacking centers (majority, up to 95% of the flat bands) and the conduction "$c$-orbitals" distributed near the AB/BA stacking centers as illustrated in Fig. 3A.

As discussed in references (*41*, *48*), the $f$-orbitals can be understood as the pseudo-Landau levels and are well localized near their centers. Therefore, the strongest Coulomb interaction $U$ and correlation effects happen among those electrons occupying the $f$-orbitals, which capture the Mott physics in MATBG. On the other hand, the topological natures of the flat bands in MATBG can be well described by the non-trivial coupling terms between the $f$- and $c$- orbitals. Such a model with both localized and itinerant orbitals strongly mimics the heavy Fermion physics, where the strongly correlated physics, such as the Kondo effect, happens at the integer filling $v_f$ of the localized orbitals, rather than the integer filling $v$ of the entire system.

At zero field, when the filling factors of these $f$-orbitals (not the entire flat bands) $v_f$ reach integers, the strong correlation effects among the $f$-electrons will greatly suppress the charge fluctuations on the $f$-orbitals, and thus minimize quasi-particle weight and compressibility near the Fermi level as indicated in Fig. 3B (assuming no additional order in the ground states). Therefore, as the function of total filling factor $v$, the local minima of compressibility should appear when the filling of the $f$-orbitals $v_f$ becomes integers, and the total filling factor $v$ will naturally be a non-integer with the deviation $\Delta v$ being the filling $v_c$ of those conduction orbitals at AB/BA stacking centers.

The above qualitative understanding is then confirmed by our numerical calculations based on the THF model for TBG using the Gutwiller variational method developed in references (*49*). As shown in Fig. 3C, the occupancy of $f$-orbitals $v_f$ shows a step-like character, while the $c$-orbitals are filled and depleted accordingly upon doping. The compressibility κ can be obtained by differentiating the occupation number respect to the total chemical potential directly (see Methods). The main results are shown in Fig. S5, where we can see the minimum compressibility indeed happens around integer $v_f$. Consequently, the experimentally observed cascade transitions can be utilized to determine the fillings of localized $f$-orbitals and itinerant $c$-orbitals. We note that in this picture, no symmetry breaking is involved to reproduce the experimentally observed "cascade transitions" (*20*, *32–36*).

Under the finite magnetic field, the appearance of CCIs is closely related to the cascade transitions at zero field discussed in the previous section. A puzzling feature in the appearance of CCIs is that for nonzero fillings only the CCIs with positive Chern numbers on the electron side and negative Chern numbers on the hole side have been detected. This can also be understood by considering the main driving force for stabilizing the symmetry-breaking states. First, the Hubbard interaction among the $f$-orbitals reaches a maximum at the integer $v_f$. Second, the CCI can be formed only when the total filling of the system $v$ satisfies the Streda formula. Therefore, the CCI will be strongly favorable when these two conditions are both satisfied as illustrated in Fig. 3D, which is only possible for the CCI with the positive (negative) Chern number on the electron (hole) side away from CNP.

**Angle-dependent phase diagram**

To further examine whether the observed link between the cascade transitions and the CCIs persists when the twist angle is slightly detuned from the magic-angle condition, we performed similar experiments on the sample with various twist angles and obtained the angle-dependent phase diagram.

Figure 4a-4c presents the measured fan diagram and the corresponding zero-field spectra for regions with twist angles of 1.10°, 1.16°, and 1.18°, respectively. The comparison of these three fan diagrams vividly illustrates the evolution of the CCIs with twist angle ($\theta$), and their onset magnetic fields $B_c$ are summarized in Fig. 4D. The regions shaded in blue represent the CCI domes, determined by the measured $B_c$ (filled triangles in blue) and the twist angle where the CCI is no longer identifiable under $B = 9$ T (intersection between filled and empty squares in blue).

In our experiments, the twist angle ranges for almost all the CCIs are larger than previous reports (*13–22*, *46*). This may be due to the sensitivity of the Rydberg sensing approach and/or the proximity-induced spin-orbit coupling effect by the adjacent $WSe_2$. Notably, the $(C, v) = (3, 1)$ state shows a minimal $B_c$ of about 0.5 T for 1.09° TBG, whereas it increases rapidly as $\theta$ is slightly detuned, resulting in the most fragile CCI state against the twist angle. In contrast, some of the CCIs could survive in a much wider range such as the $(\pm 1, \pm 3)$ and $(-2, -2)$ states, which are still observable up to 1.24° at $B = 9$ T, significantly beyond the magic angle. Moreover, one might intuitively expect that the stabilization of the CCIs needs a larger $B_c$ when the twist angle deviates from the magic angle. However, the experimental data reveals a more intricate relationship as exemplified by the $(-2, -2)$ state, where the $B_c$ gets minimized when the twist angle is around 1.18°. These behaviors could stem from the multiple competing orders within the system, where minor variations in experimental conditions may prompt a shift to a different energetically favorable ground state.

We then turn our attention to the link between the cascade transitions and the CCIs. As depicted in Fig. 4A to 4C, for almost all the zero-field peaks in the lower panel, its filling factor aligns with the onset of CCIs. Another trend is that, at larger twist angles, some CCIs may not necessarily originate from a zero-field peak, such as the $(C, v) = (\pm 1, \pm 3)$ state for $\theta = 1.18°$. The peak feature is absent around $v = \pm 3$ in the lower panel while the corresponding CCIs still emerge at high magnetic fields.

In Fig. 4d, we also plot the zero-field deviation $\Delta v$ multiplied by a constant $\Phi_0/C$ for samples with various twist angles near different integer fillings with red triangles. The red dashed lines denote the largest angles where the cascade transitions exist, demonstrating their relatively narrower ranges of the twist angle compared to the CCIs. In such angle ranges, the values of $B_c$ and $\Phi_0\Delta v/C$ exhibit a concurrence. It indicates the resilience of the relationship $B_c = \Phi_0\Delta v/C$ and the proposed mechanism across a broad spectrum of twist angles.

To summarize, we resolve the cascade transitions and CCIs in MATBG through Rydberg exciton sensing and reveal their previously hidden connection which could be explained by the THF model. The cascade transitions manifested by asymmetric sawtooth features stem from the doping-dependent redistribution of charges between the localized $f$- and itinerant $c$- orbitals that does not require any symmetry breaking. Notably, both compressibility minima in the cascade transitions and the onsets of CCIs occur at integer fillings of the localized $f$-orbitals $v_f$, rather than integer fillings of the entire system $v$. Many puzzling experimental observations can now be well explained within this framework. For example, the finite $B_c$ of CCIs is the natural result of the finite $\Delta v$ contributed by topological itinerant $c$ electrons, hindering the observations of the QAHE in non-hBN aligned MATBG. Meanwhile, the non-diverging resistance peaks and the Pomeranchuk effect, which are both commonly found to occur at non-integer fillings of the entire system, are likely to arise from the suppression of the quasiparticle weight at integer $v_f$ and local moment fluctuations of $f$-orbitals at higher temperatures (*21*, *34*, *35*).

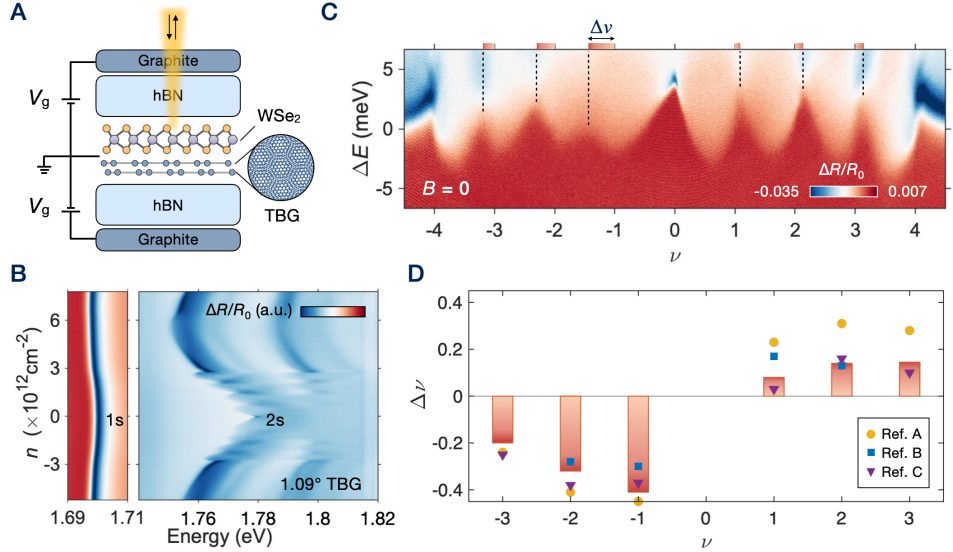

**Fig. 1. Optical detection of the zero-field compressibility evolution in 1.09° TBG. (A)** Schematic illustration of the device structure and optical measurement. The TBG-WSe$_2$ is dual-gated by graphite-hBN dielectrics where the gate voltage $V_g$ is applied. Optical detection of the compressibility evolution is facilitated through the Rydberg sensing scheme utilizing WSe$_2$ 2s excitons. **(B)** Doping dependence of the reflectance contrast spectra of the device. The sawtooth feature in the 2s state originates from the nearly periodic changes in the electronic compressibility of TBG. **(C)** Magnified view of the sawtooth feature in the coordinate of $\Delta E$ (energy relative to a smooth baseline) and $\nu$ (moiré filling factor). The peaks at $\nu = \pm 4$ align with the band insulator states, while those around $\nu = \pm 1, \pm 2, \pm 3$ signify the other six compressibility minima that correspond to the cascade transitions. The red bars denote the deviation $\Delta\nu$ of the compressibility minima (indicated by dashed lines) from nearby integer fillings. **(D)** Deviation of the compressibility minima $\Delta\nu$ extracted from our work (bar) and previous compressibility measurements (symbols). Refs. A, B, and C are adapted from scanning electron transistor (*32*), quantum capacitance (*22*), and direct chemical potential (*34*) measurements, respectively.

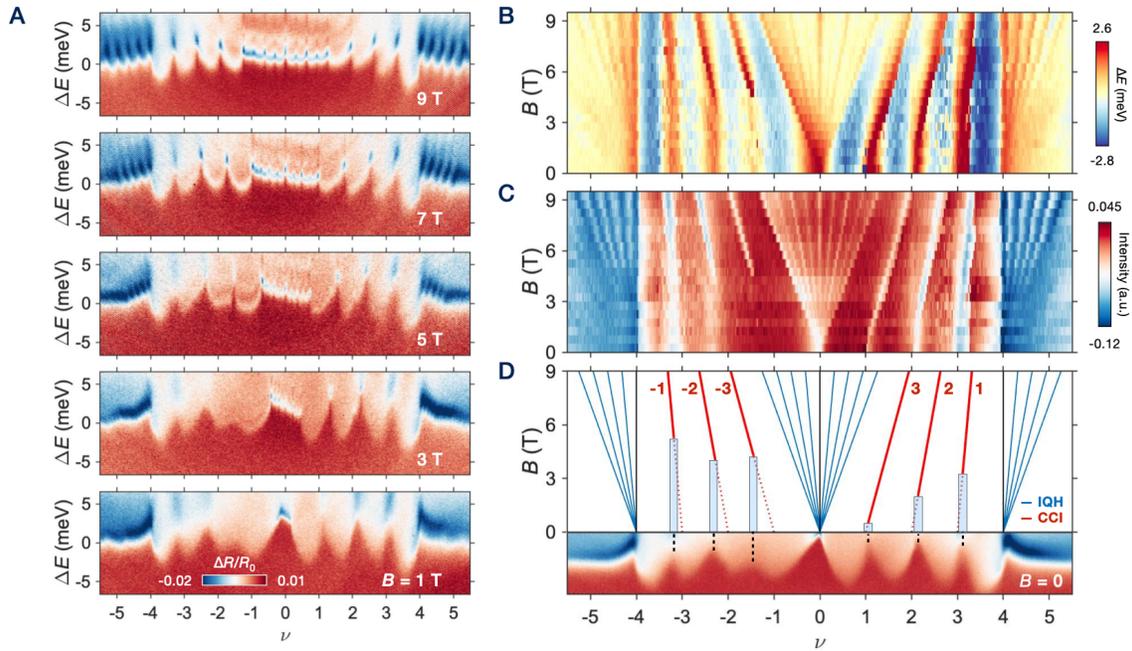

**Fig. 2. Correlated Chern insulators (CCIs) emerging at finite magnetic fields. (A)** The spectral evolution of the sample under a perpendicular magnetic field. With increasing field strength, more peaks become identifiable and gradually shift in doping densities. **(B) (C)** Fan diagram constructed with the extracted energy $\Delta E$ (B) and intensity (C) of the 2s state. Features with large $\Delta E$ and small intensity are associated with discernible states. **(D)** Schematics of the fan diagram (upper) and the spectrum in Fig. 1c for reference (lower). The blue lines denote the Landau fan sequences originating from $\nu = 0$ and $\nu = \pm 4$. The red lines signify the CCIs emanating from $\nu = \pm 1, \pm 2, \pm 3$ (guided by the dashed lines), and their corresponding Chern numbers are marked in red. As indicated by the blue bars, the onsets of the CCIs consistently occur at the doping density where the electronic compressibility is minimized at zero magnetic fields (marked by the vertical dashed lines).

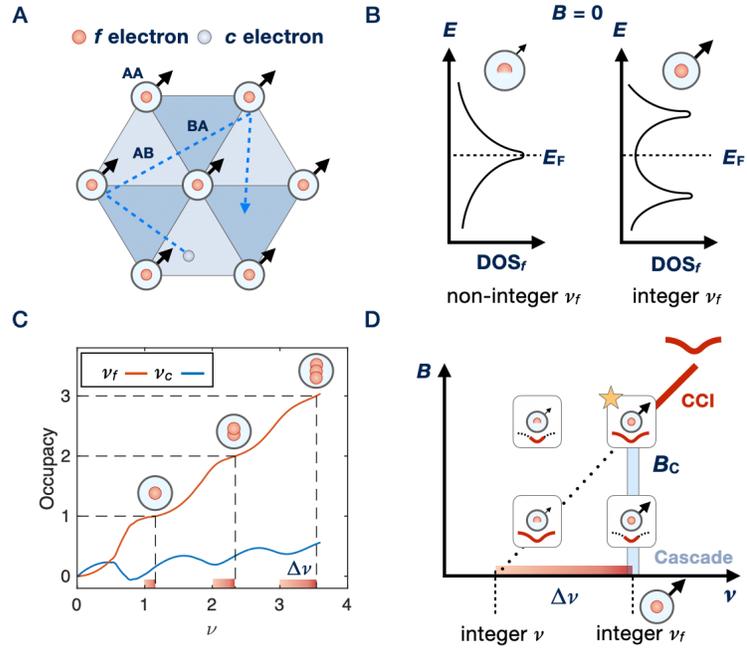

**Fig. 3. Topological heavy fermion (THF) picture for understanding the link between cascade transitions and CCIs.** (**A**) Schematic illustration of the THF model with localized *f*-electrons (centered at AA stacking sites) and itinerant *c* electrons (centered at AB/BA stacking sites). (**B**) Schematics of the density of state suppression at integer $v_f$. When the *f*-orbitals are occupied by integer electrons/holes, the suppression of the charge fluctuation on the *f*-orbitals and the electronic correlation becomes strongest, leading to a minimum in local electronic density of state (DOS). (**C**) Calculated occupancy of the *f*- and *c*- orbitals upon electron filling. The integer filling of the *f*-orbitals $v_f$ naturally occurs at non-integer $v$, and the deviation $\Delta v$ equals the filling of the conduction orbitals $v_c$. (**D**) Schematics of the emergence of a CCI state exemplified by the electron-doped side. Stabilizing CCI requires a strong Hubbard interaction among the *f*-orbitals and a total filling $v$ satisfying the Streda formula. As a result, the CCIs emerge when the integer $v_f$ (blue bar) crosses the Streda line (dotted line), accounting for the observed link between cascade transitions and CCIs.

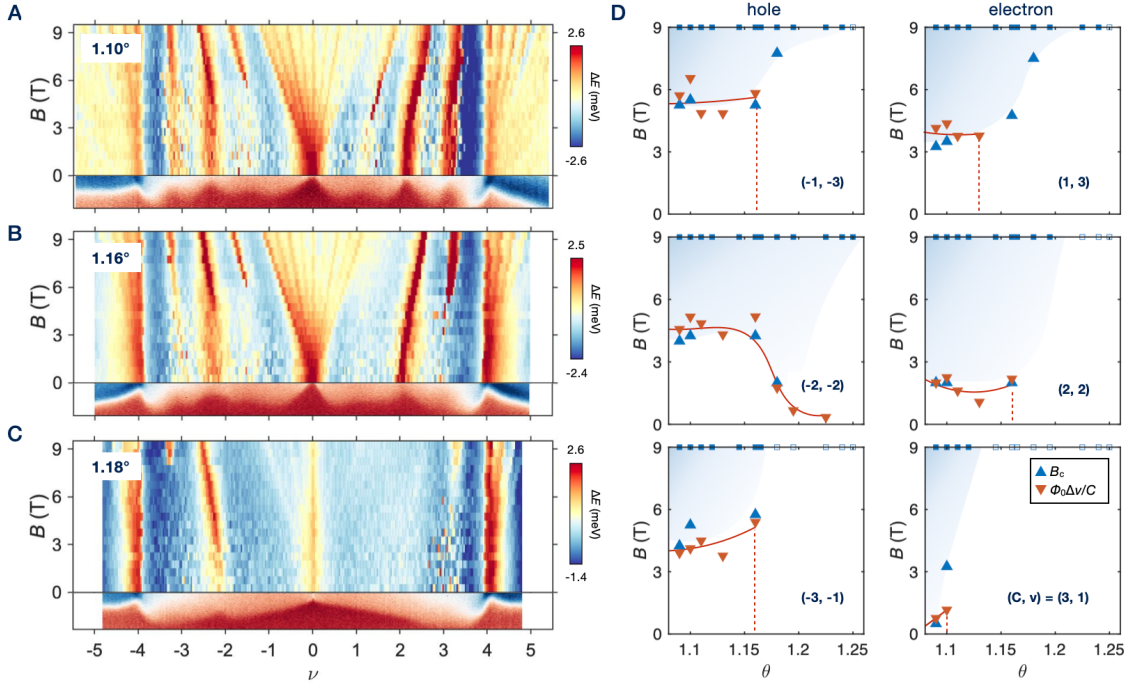

**Fig. 4. Angle-dependent phase diagram of the CCIs and their links with the zero-field spectra.** **(A), (B) and (C),** Fan diagram (upper) and spectra at $B=0$ (lower) of the samples with twist angles 1.10° (A), 1.16° (B), and 1.18° (C). **(D)** CCI phase diagram highlighted in the shadowed blue regions defined by the critical magnetic field $B_c$ (triangles in blue) and the critical twist angle at $B = 9$ T. The critical twist angle is determined by the boundary between filled squares (existence of CCIs) and empty squares (absence of CCIs) in blue. The dashed red lines denote the boundary in the twist angle of cascade features. The red triangles indicate the deviation of the zero-field compressibility minimum from integer fillings $\Delta v$ multiplied by $\Phi_0/C$. The trend is guided by the red curves, showing a close correspondence with the critical magnetic field $B_c$ over a broader range detuned from the magic angle.

# Supplementary Materials for
## Link between cascade transitions and correlated Chern insulators in magic-angle twisted bilayer graphene

Qianying Hu[1,2,#], Shu Liang[3,#], Xinheng Li[2], Hao Shi[3], Xi Dai[3,*], Yang Xu[1,2,*]

[1]Beijing National Laboratory for Condensed Matter Physics, Institute of Physics, Chinese Academy of Sciences, Beijing, China
[2]School of Physical Sciences, University of Chinese Academy of Sciences, Beijing, China
[3]Department of Physics, Hong Kong University of Science and Technology, Kowloon, Hong Kong, China

[#]These authors contributed equally to this work.
[*]Email: daix@ust.hk, yang.xu@iphy.ac.cn

## Methods

**Device fabrication and electrostatic gating**

The device fabrication process is identical to that described in our previous work (*39*) using the standard dry-transfer (*50*) and tear-and-stack (*51*) methods. The WSe$_2$, hBN, graphene, and few-layer graphite are mechanically exfoliated from bulk crystals and picked up layer by layer. The stack is released onto silicon substrates with pre-patterned gold electrodes where the gate voltage $V_g$ is symmetrically applied by Keithley 2400 source meters. The carrier density in TBG is calibrated by the spacing $n_0 = eB/h$ of the Landau level spectral features, where $e$ is the elementary charge. The twist angle is calculated from the full-filling density $n_s$ at superlattice filling factor $v = \pm 4$ through $n_s \approx 8\theta^2/\sqrt{3}a^2$ (the graphene lattice constant $a = 0.246$ nm).

**Reflectance spectroscopy measurements**

The devices are measured in a close-cycle cryostat attoDry2100 at a base temperature of 1.7 K under a perpendicular magnetic field up to 9 T. The light source is a halogen lamp whose output is collected by a single-mode fiber and collimated by a ×10 objective lens. Magneto-optical measurements are performed with a circularly polarized light created by a linear polarizer and a quarter wave plate. A low-temperature compatible objective (NA=0.82) focuses the beam onto the sample with a diameter of ~ 1 μm and a power lower than a few nW. The reflectance from the sample is collected by the same objective and detected by a spectrometer with 600 lines/mm gratings for fan-diagram measurements and 1800 lines/mm gratings for fine gate-dependent spectra measurements.

**Background removing and extraction of energy and intensity**

The optical measurement of TBG in this paper is based on the doping-dependent optical resonance of the 2s Rydberg exciton. The reflectance contrast ($\Delta R/R_0$) spectrum is obtained by comparing the spectrum from the sample ($R$) with that from the substrate immediately next to the sample ($R_0$) as $\Delta R/R_0 = (R - R_0)/R_0$. The compressibility evolution of TBG will affect its energy and intensity in the gate-dependent $\Delta R/R_0$ spectrum. We aim to extract these two values properly and construct fan diagrams.

The raw data of the 1.09° device under $B = 0$ and $B = 9$ T are plotted in Fig.s S1 A and D, respectively. We first subtract the background noise for more precise energy extraction in the next step (method described in ref (*39*)). By extracting the minima of $d(\Delta R/R_0)/dE$, we obtain and plot the absolute 2s resonance energy in blue in these two figures. However, the absolute energy of the 2s exciton upon doping is also affected by other issues beyond TBG compressibility, i.e. the spatial confinement of the exciton wave function by TBG moiré potential (formation of Rydberg moiré exciton). This effect is prominent in small-angle TBG devices as discussed in our previous work (*39*). It is still not negligible in MATBG, manifesting as the non-monotonous energy shift upon doping in Fig. 1B.

To isolate the contribution from TBG compressibility evolution, we focus on the relative energy ($\Delta E$) to the featureless baseline obtained by smoothing the extracted 2s energy in this paper. As shown in Fig.s S1 B and E, the spectra after background subtraction are presented in the coordinate of $\Delta E$. This region is between the yellow (7 meV lower than the baseline) and orange (7 meV higher than the baseline) curves in Fig. S1A and D. The relative energy $\Delta E$ of the 2s exciton (blue curves) is then used in the magnetic field-dependent counterplot in Fig. 2B. We extract the spectral intensity by integrating the $\Delta R/R_0$ in the region defined by blue and red lines in Fig. S1B and E (~2 meV energy span). The integrated intensity is shown in Fig. S1C and F, which is used to plot the fan diagram in Fig. 2C after normalization.

**Method for calculation of no symmetry-breaking state:**

**Single Particle TB-model:** The single particle Hamiltonian of TBG in each valley and spin can be described by 8 orbitals(*52*) (real-space distribution illustrated in Fig. S4). They consist of two p± orbitals localized at AA point (f1: $p+$ @AA, f2: $p-$ @AA), a ring-shaped s orbital centered at AA point (c1: s @AA), a pair of $p_z$ orbitals localized at AB and BA points (c2: $p_z$ @AB, c3: $p_z$ @BA), and three s orbitals localized at the three domain walls (c4: s @DW1, c5: s @DW2, c6: s @DW3). The first 2 orbitals are strongly correlated *f* orbitals (95% of flat bands), while the other 6 orbitals are weakly correlated *c* orbitals (5% of flat bands).

When considering the valley and spin degeneracy, we have $32 = 8(f) + 24(c)$ orbitals in total per unit cell. The effective TB Hamiltonian reads:

$$\hat{H}_0^{s\eta}(\mathbf{k}) = \sum_{tt'}^{1\sim 2} \left[\mathcal{H}_{ff}^{\eta}(\mathbf{k})\right]_{tt'} f_{\mathbf{k}t\eta s}^{\dagger} f_{\mathbf{k}t'\eta s} + \sum_{aa'}^{1\sim 6} \left[\mathcal{H}_{cc}^{\eta}(\mathbf{k})\right]_{aa'} c_{\mathbf{k}a\eta s}^{\dagger} c_{\mathbf{k}a'\eta s}$$

$$+ \sum_{t}^{1\sim 2} \sum_{a}^{1\sim 6} \left(\left[\mathcal{H}_{fc}^{\eta}(k)\right]_{ta} f_{\mathbf{k}t\eta s}^{\dagger} c_{\mathbf{k}a\eta s} + h.c.\right)$$

$$\hat{H}_0 = \sum_{\mathbf{k}s\eta} \hat{H}_0^{s\eta}(\mathbf{k})$$

Where $s$ and $\eta$ denote spin and valley, $t$ and $a$ denote "*f*" and "*c*" orbitals respectively.

**Model Hamiltonian:** The total Hamiltonian is written as:

$$\hat{H} = \hat{H}_0 + \hat{U} + \hat{W} + \hat{V}$$

$$\hat{U} = \frac{U}{2} \sum_{R} \sum_{\alpha \neq \beta} (\hat{n}_{Rf\alpha} - n_{Rf\alpha}^{CNP})(\hat{n}_{Rf\beta} - n_{Rf\beta}^{CNP})$$

$$\widehat{W} = W \sum_{R} \sum_{\alpha\beta} (\hat{n}_{Rf\alpha} - n_{Rf\alpha}^{CNP})(\hat{n}_{Rc\beta} - n_{Rc\beta}^{CNP})$$

$$\widehat{V} = \frac{V}{2} \sum_{R} \sum_{\alpha \neq \beta} (\hat{n}_{Rc\alpha} - n_{Rc\alpha}^{CNP})(\hat{n}_{Rc\beta} - n_{Rc\beta}^{CNP})$$

Here $\widehat{U}, \widehat{W}, \widehat{V}$ are the onsite interaction among "f-orbitals", effective interaction between "f-orbitals" and "c-orbitals", and effective interaction among "c-orbitals", respectively. The $\widehat{U}$ is treated with Gutzwiller approximation while $\widehat{W}, \widehat{V}$ are treated at Hartree level in our calculation (specified below). The α, β are combined indices of spin, valley, and orbitals. The $R$ marks the moiré unit cell. Double countings are considered within $\widehat{U}, \widehat{W}, \widehat{V}$ by subsracting density operator with its occupation value at charge neutrality point (CNP). The interaction parameters used in this work are: $U = 0.08$ eV, $W = 0.07$ eV, $V = 0.07$ eV. Here $\widehat{W}, \widehat{V}$ are effective interactions in the sense that they are an approximation of realistic nearest neighbor interaction involving "c-orbitals" which makes $W$ and $V$ become many times larger than actually calculated values(36). In addition, the effective $\widehat{W}, \widehat{V}$ won't break any symmetry here since they're treated at Hartree level. We adopt such approximation because we identify $\widehat{U}$ as our primary contribution to the correlated effect in this system and we would like to reduce the total number of parameters.

**Gutzwiller Variational Scheme:** First we introduce the Gutzwiller correlator(49, 53, 54) for site $\boldsymbol{R}$:

$$\hat{P}_{\boldsymbol{R}} = \sum_{I} \lambda_{\boldsymbol{R};I} |\boldsymbol{R};I\rangle\langle\boldsymbol{R};I|$$

Where $\lambda_{\boldsymbol{R};I}$ is the variational parameter, $|\boldsymbol{R};I\rangle$ is many-body Fock state consisting of local "f-orbitals" and $|\boldsymbol{R};I\rangle\langle\boldsymbol{R};I|$ is a projector which projects out states other than $|\boldsymbol{R};I\rangle$ (We'll use α as integrated indices for $s \otimes \eta \otimes t$ in this section):

$$|\boldsymbol{R};I\rangle = \prod_{\alpha \in I} \hat{f}_{\boldsymbol{R};\alpha}^{\dagger} |vaccum\rangle$$

$$|\boldsymbol{R};I\rangle\langle\boldsymbol{R};I| = \prod_{\alpha \in I} \hat{n}_{\boldsymbol{R};f\alpha} \prod_{\beta \notin I} (1 - \hat{n}_{\boldsymbol{R};f\beta})$$

Here we use a diagonal $\hat{P}_{\boldsymbol{R}}$ because only density-density interaction is included. The Gutzwiller variational wavefunction is obtained by acting $\hat{P}_{\boldsymbol{R}}$ on a trial non-interacting Fermi Sea $|\Phi_0\rangle$:

$$|\Psi_G\rangle = \prod_{R} \hat{P}_{\boldsymbol{R}} |\Phi_0\rangle$$

Gutzwiller approximation puts some constraints on $\lambda_I$:(53, 55)
$$\langle \Phi_0 | \hat{P}_R^2 | \Phi_0 \rangle = 1$$
$$\langle \Phi_0 | \hat{P}_R^2 \hat{n}_{Rf\alpha} | \Phi_0 \rangle = \langle \Phi_0 | \hat{n}_{Rf\alpha} | \Phi_0 \rangle$$

The total energy is then ($N_m$: number of unit cells or k points):

$$E_{total} = \frac{1}{N_m} \langle \Psi_G | \widehat{H}_0 + \widehat{U} + \widehat{W} + \widehat{V} | \Psi_G \rangle$$

The first term is the kinetic energy (we denote $\langle \hat{O} \rangle_0 = \langle \Phi_0 | \hat{O} | \Phi_0 \rangle$):

$$E_{kin} = \frac{1}{N_m} \sum_{ks\eta} \sum_{tt'}^{1\sim 2} \left[\mathcal{H}_{ff}^{\eta}(\boldsymbol{k})\right]_{tt'} \mathcal{Z}_{t\eta s} \langle f_{\boldsymbol{k}t\eta s}^{\dagger} f_{\boldsymbol{k}t'\eta s} \rangle_0 \mathcal{Z}_{t'\eta s} + \frac{1}{N_m} \sum_{ks\eta} \sum_{aa'}^{1\sim 6} \left[\mathcal{H}_{cc}^{\eta}(\boldsymbol{k})\right]_{aa'} \langle c_{\boldsymbol{k}a\eta s}^{\dagger} c_{\boldsymbol{k}a'\eta s} \rangle_0$$

$$+ \sum_{t}^{1\sim 2} \sum_{a}^{1\sim 6} \left( \left[\mathcal{H}_{fc}^{\eta}(\boldsymbol{k})\right]_{ta} \mathcal{Z}_{t\eta s} \langle f_{\boldsymbol{k}t\eta s}^{\dagger} c_{\boldsymbol{k}a\eta s} \rangle_0 + h.c. \right)$$

Where $\mathcal{Z}_{\alpha}$ is the renormalization factor (the site index $\boldsymbol{R}$ is omitted from now; $n_{f\alpha}^0 = \langle \hat{n}_{Rf\alpha} \rangle_0$):

$$\mathcal{Z}_{\alpha} = \lambda^{\mathsf{T}} \mathbb{R}_{\alpha} \lambda$$

$$(\mathbb{R}_\alpha)_{II'} = \frac{1}{2}\sqrt{\frac{m_I^0 m_{I'}^0}{n_{f\alpha}^0(1-n_{f\alpha}^0)}}(\delta_{I,I'\cup\alpha} + \delta_{I\cup\alpha,I'})$$

$$m_I^0 = \prod_{\alpha\in I} n_{f\alpha}^0 \prod_{\beta\notin I}(1-n_{f\beta}^0)$$

The second term is the interaction energy among "$f$-orbitals":

$$E_{int}^{ff} = \lambda^\mathsf{T}\mathbb{H}_{int}\mathbb{m}^0\lambda - U\sum_{\alpha\neq\beta} n_{f\alpha}^0 n_{f\beta}^{CNP} + \frac{U}{2}\sum_{\alpha\neq\beta} n_{f\alpha}^{CNP} n_{f\beta}^{CNP}$$

$$(\mathbb{H}_{int})_{II'} = \delta_{II'}\frac{U}{2}\langle I|\sum_{\alpha\neq\beta}\hat{n}_{f\alpha}\hat{n}_{f\beta}|I\rangle$$

$$(\mathbb{m}^0)_{II'} = \delta_{II'} m_I^0$$

Since $\hat{P}_R$ does not contain "$c$-orbitals", $\langle\Psi_G|\widehat{W}+\widehat{V}|\Psi_G\rangle = \langle\Phi_0|\widehat{W}+\widehat{V}|\Phi_0\rangle$ under Gutzwiller approximation. Using Hartree approximation for these 2 terms:

$$\langle\Phi_0|\widehat{W}|\Phi_0\rangle = W(n_f^0 - n_f^{CNP})(n_c^0 - n_c^{CNP})$$

$$\langle\Phi_0|\widehat{V}|\Phi_0\rangle \approx \frac{V}{2}(n_c^0 - n_c^{CNP})(n_c^0 - n_c^{CNP})$$

Here we denote $n_f^0 = \sum_\alpha n_{f\alpha}^0$ and $n_c^0 = \sum_\beta n_{c\beta}^0$. Notice that we dropped the self-energy in $\langle\Phi_0|\widehat{V}|\Phi_0\rangle$ since we want to reduce the number of variables in the final energy functional (using the total occupancy of "$c$-orbitals" instead of their individual occupancies, which will become clear in the following text).

We use an efficient variational scheme developed in ref.(*49*) where orbital occupancies are input variables for the energy functional. Since we're considering a no-symmetry breaking state, only the total occupancy $n_f$ (denoted as $v_f$ in the main text) matters for $f$-orbitals as $n_{f\alpha} = \frac{n_f}{8}$. To reduce the number of total variables in the energy functional, we only consider the total occupancy for $c$-orbitals $n_c$ (denoted as $v_c$ in the main text). The energy functional $E_G = E_{total}$ when variational constraints are satisfied:

$$E_G(n_f, n_c) = \frac{1}{N_m}\langle\Psi_G|\widehat{H}_0 + \widehat{U}|\Psi_G\rangle + W(n_f - n_f^{CNP})(n_c - n_c^{CNP}) + \frac{V}{2}(n_c - n_c^{CNP})(n_c - n_c^{CNP})$$

$$+ \sum_\alpha \lambda_{f\alpha}^\mathrm{F}(\langle\hat{n}_{Rf\alpha}\rangle_0 - \frac{n_f}{8}) + \lambda_c^\mathrm{F}(\sum_\beta\langle\hat{n}_{Rc\beta}\rangle_0 - n_c) + \sum_\alpha \lambda_{f\alpha}^\mathrm{B}(\sum_{\alpha\in I}\lambda_I^2 m_I^0$$

$$- \frac{n_f}{8}) + E^F(1 - \langle\Phi_0|\Phi_0\rangle) + E^B(1 - \sum_I \lambda_I^2 m_I^0)$$

Where $\lambda^F, \lambda^B, E^F, E^B$ are Lagrange multipliers for constraints on $|\Phi_0\rangle$ and $\boldsymbol{\lambda}$. Given a set of $(n_f, n_c)$, we treat $|\Phi_0\rangle$ and $\boldsymbol{\lambda}$ as independent variables which should minimize $E_G|_{n_f,n_c}$:

$$\frac{\delta E_G}{\delta\langle\Phi_0|}\Big|_{n_f,n_c,\lambda} = 0 \rightarrow \widehat{H}^F|\Phi_0\rangle = \mathrm{E}^F|\Phi_0\rangle$$

(1)

$$\frac{\delta E_G}{\delta\lambda}\Big|_{n_f,n_c,|\Phi_0\rangle} = 0 \rightarrow \widehat{H}^B\lambda = E^B\mathbb{m}^0\lambda$$

(2)

$$\widehat{H}^F = \sum_{ks\eta} \left\{ \sum_{tt'}^{1\sim 2} \left[\mathcal{H}^\eta_{ff}(\boldsymbol{k})\right]_{tt'} \mathcal{Z}_{t\eta s} f^\dagger_{\boldsymbol{k}t\eta s} f_{\boldsymbol{k}t'\eta s} \mathcal{Z}_{t'\eta s} + \sum_{aa'}^{1\sim 6} \left[\mathcal{H}^\eta_{cc}(\boldsymbol{k})\right]_{aa'} c^\dagger_{\boldsymbol{k}a\eta s} c_{\boldsymbol{k}a'\eta s} \right.$$

$$+ \sum_t^{1\sim 2} \sum_a^{1\sim 6} \left( \left[\mathcal{H}^\eta_{fc}(\boldsymbol{k})\right]_{ta} \mathcal{Z}_{t\eta s} f^\dagger_{\boldsymbol{k}t\eta s} c_{\boldsymbol{k}a\eta s} + h.c. \right) + \sum_t \lambda^F_{f;t\eta s} \hat{n}_{f;\boldsymbol{k}t\eta s}$$

$$\left. + \lambda^F_c \left( \sum_a^{1\sim 6} \hat{n}_{c;\boldsymbol{k}a\eta s} \right) \right\}$$

$$\widehat{H}^B = \sum_\alpha \chi_\alpha \mathbb{R}_\alpha + \mathbb{H}_{int} \mathbb{m}^0 + \sum_\alpha \lambda^B_{f\alpha} \mathbb{N}_\alpha \mathbb{m}^0$$

$$\chi_\alpha = \frac{2}{N_m} \sum_{\boldsymbol{k}s\eta} \sum_{tt'}^{1\sim 2} \left[\mathcal{H}^\eta_{ff}(\boldsymbol{k})\right]_{tt'} \langle f^\dagger_{\boldsymbol{k}t\eta s} f_{\boldsymbol{k}t'\eta s} \rangle_0 \mathcal{Z}_{t'\eta s} + \sum_t^{1\sim 2} \sum_a^{1\sim 6} \left( \left[\mathcal{H}^\eta_{fc}(\boldsymbol{k})\right]_{ta} \langle f^\dagger_{\boldsymbol{k}t\eta s} c_{\boldsymbol{k}a\eta s} \rangle_0 + h.c. \right)$$

$$(\mathbb{N}_\alpha)_{II'} = \langle I | \hat{n}_{f\alpha} | I' \rangle$$

$min\, E_G|_{n_f, n_c}$ can be found by solving the 2 linear equations (1) and (2) self-consistently. The solution of (1) gives us $\chi_\alpha$ which serves as inputs for (2). Conversely, the solution of (2) gives us $\mathcal{Z}_\alpha$ which serves as inputs for (1).

Given an electron filling $\nu$, we have $n_f + n_c = \nu$. The inverse compressibility $\kappa^{-1} = \frac{\partial^2 min\, E_G}{\partial \nu^2}$ is calculated numerically. The presented result for inverse compressibility is smoothed to compensate numerical errors.

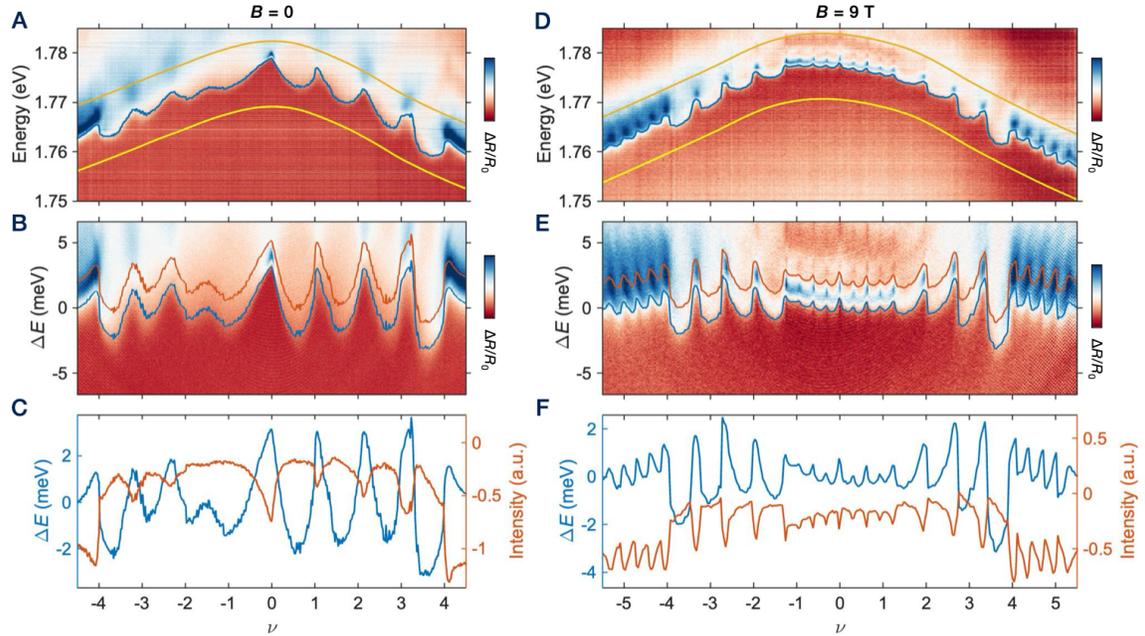

**Fig. S1. Background removing and extraction of the energy and intensity. (A) (D)** Raw data of the 1.09° device under $B = 0$ and $B = 9$ T. The blue curves are the extracted 2s resonance energy. **(B) (E)** Data after background subtraction in the coordinate of $\Delta E$ (corresponding to the region between yellow and orange curves in (A) and (D). **(C) (F)** Extracted relative energy $\Delta E$ (in blue) and the integrated intensity between red and blue curves in (B) and (E) (in red).

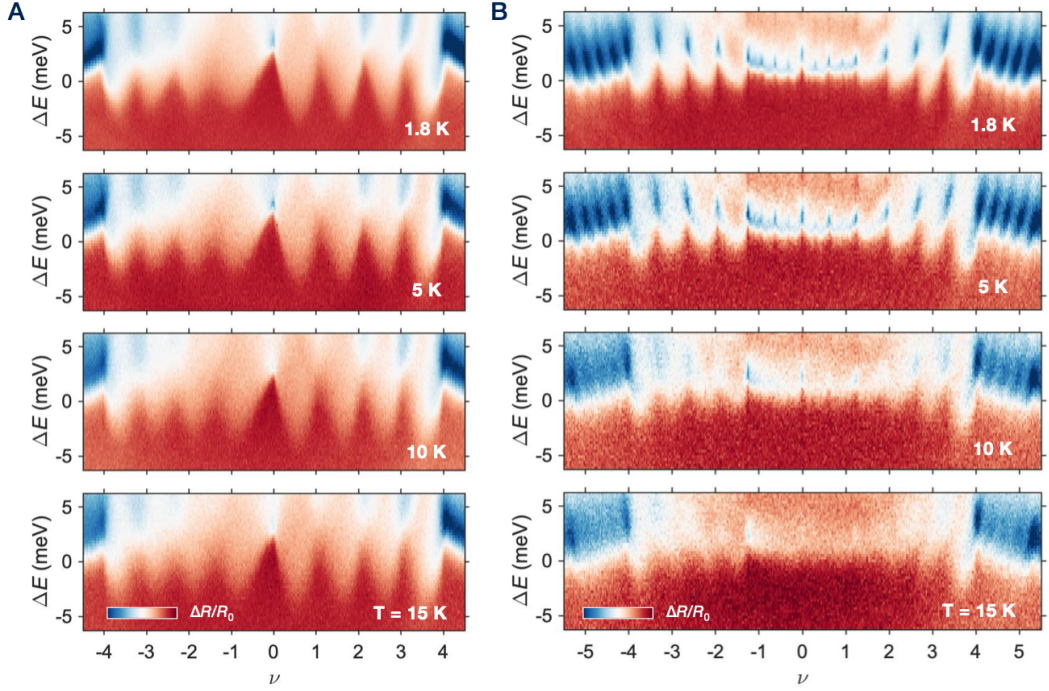

**Fig. S2. Temperature dependence of the cascade transitions and CCIs. (A)** Reflectance contrast spectra of the 1.09° device at $T$ = 1.8 K, 5 K, 10 K, and 15 K. The sawtooth feature mainly remains unchanged with increasing temperature up to ~15 K. **(B)** Reflectance contrast spectra of the 1.09° device at $T$ = 1.8 K, 5 K, 10 K, and 15 K under $B$ = 9 T. The blue tails of the insulating states sequentially disappear upon heating, suggesting the thermal activation behavior of the CCI gaps.

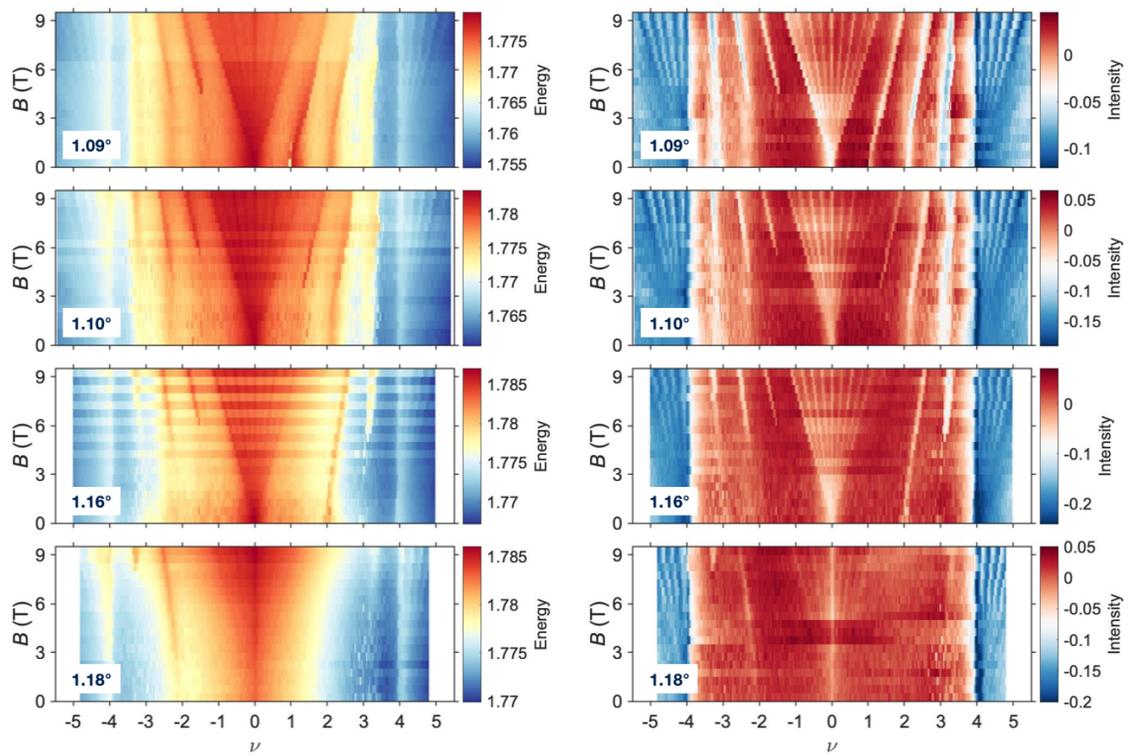

**Fig. S3. Additional fan diagram of the 1.09°, 1.10°, 1.16°, and 1.18° device plotted with 2s energy (left) and integrated intensity (right).**

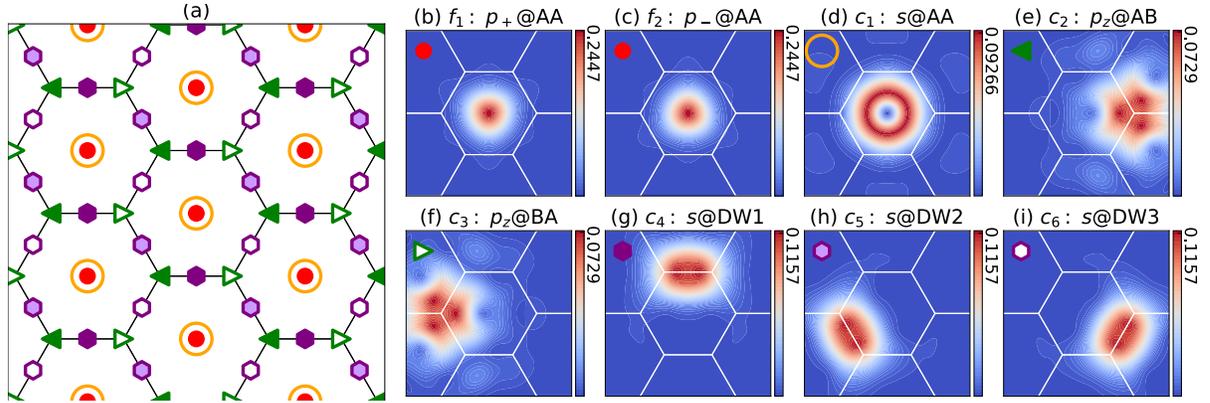

**Fig. S4. Real-space distribution of the 8 orbitals. (A)** The lattice of the 8-band model in $\eta = +$ valley. Different symbols represent different sublattices. **(B)-(I)** The density of these orbitals in the first cell.

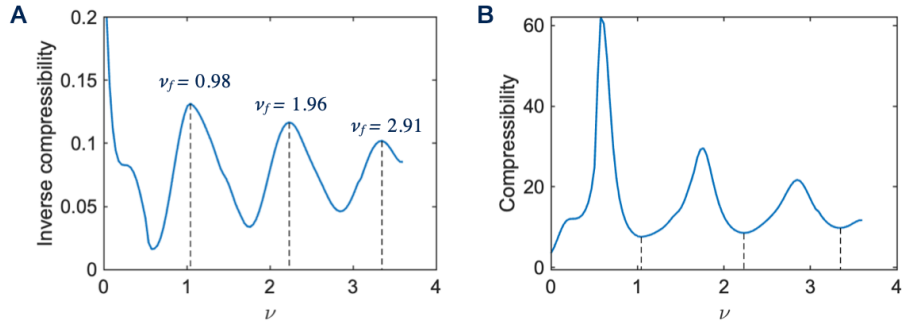

**Fig. S5. Calculated inverse compressibility $\partial\mu/\partial\nu$ (A) and compressibility $\partial\nu/\partial\mu$ (B) based on the THF model.** In such non-symmetry breaking states, only suppression (non-zero) of quasiparticle weight occurs near the integer fillings of localized *f* orbitals.
24